\documentclass[conference]{IEEEtran}
\IEEEoverridecommandlockouts
\usepackage{amsmath,amssymb,amsfonts}
\usepackage{xcolor}
\usepackage{caption,subcaption}
\usepackage{standalone}
\usepackage{url}
\usepackage{subcaption}
\usepackage{algorithm,algorithmicx,algpseudocode}
\usepackage{rotating}
\usepackage{tikz}
\usetikzlibrary{fit,graphs,backgrounds,scopes}
\usepackage{pgf}
\usepackage{pgfplots,gincltex}
\usepackage{svg}
\usepackage{booktabs}
\usetikzlibrary{calc}
\usetikzlibrary{positioning}
\usetikzlibrary{matrix}
\usepackage{multirow}
\usepackage{listings}
\usepackage{pifont}
\usepackage{todonotes}
\usepackage[capitalize]{cleveref}
\crefname{paragraph}{paragraph}{paragraphs}
\Crefname{paragraph}{Paragraph}{Paragraphs}

\newcommand{\intextstyle}{\ttfamily\small}

    {\noindent\ignorespaces\begin{lstlisting}[#1,float,floatplacement=H]}{\end{lstlisting}\noindent\ignorespacesafterend}

\definecolor[named]{whitesmoke}   {rgb}{0.96,0.96,0.96}
\definecolor[named]{codegreen}    {cmyk}{0.20,0,1,0.19}
\definecolor[named]{codered}      {cmyk}{0,0.90,0.86,0}
\definecolor[named]{codelightblue}{cmyk}{0.49,0.01,0,0}
\definecolor[named]{codedarkblue} {cmyk}{1,0.58,0,0.21}
\makeatletter
\newenvironment{btHighlight}[1][]
{\begingroup\tikzset{bt@Highlight@par/.style={#1}}\begin{lrbox}{\@tempboxa}}
{\end{lrbox}\bt@HL@box[bt@Highlight@par]{\@tempboxa}\endgroup}

\newcommand\btHL[1][]{%
    \begin{btHighlight}[#1]\bgroup\aftergroup\bt@HL@endenv%
}
\def\bt@HL@endenv{%
    \end{btHighlight}%
    \egroup
}
\newcommand{\bt@HL@box}[2][]{%
  \ifdefined\tikzexternaldisable%
    \tikzexternaldisable%
  \fi%
    \tikz[#1]{%
        \pgfpathrectangle{\pgfpoint{1pt}{0pt}}{\pgfpoint{\wd #2}{\ht #2}}%
        \pgfusepath{use as bounding box}%
        \node[anchor=base west, fill=green,outer sep=0pt,inner xsep=1pt, inner ysep=0pt, rounded corners=3pt, minimum height=\ht\strutbox+1pt,#1]{\raisebox{1pt}{\strut}\strut\usebox{#2}};
    }%
  \ifdefined\tikzexternalenable%
    \tikzexternalenable%
  \fi%
}

\lstdefinestyle{node}{
    backgroundcolor=,
    language=,
    basicstyle=\tiny\ttfamily,
    morekeywords = {br,neg,or,and,all,any},
    numbers=none,
    mathescape=true,
    frame=none,
    literate={<-}{{$\leftarrow$}}1
}

\lstdefinelanguage{alg}{
    morecomment = [s]{/*}{*/},
    morecomment = [l]{//},
    sensitive = true,
    morekeywords = {for,next,to,step}
}

\lstdefinelanguage{impala}{
    morecomment = [s]{/*}{*/},
    morecomment = [l]{//},
    sensitive = true,
    morekeywords = {i8,i16,i32,i64,u8,u16,u32,u64,f16,f32,f64,bool,int,float,double,extern,struct,as,match,true,false,type,with,let,mut,static,while,in,exit,return,break,continue,if,else,for,do,fn,any,all,extract,shuffle,ballot,enum,stream},
    morekeywords = {@,@@},
    morestring = [b]",
    moredelim=**[is][\btHL]{§}{§},
}

\lstdefinelanguage{metaocaml}{
    sensitive = true,
    morekeywords = {let,in,rec,if,then,else,fun},
    moredelim=**[is][\btHL]{§}{§},
}

\lstdefinelanguage{pseudoml}{
    sensitive = true,
    morekeywords={fun,where,whererec,lambda,let,letrec,in,and,bool,float,int,br,noret},
    literate=%
        {==}{{=}}1
        {!=}{{$\neq$}}1
        {<=}{{$\leq$}}1
        {>=}{{$\geq$}}1
        {->}{{$\rightarrow$}}1
        {<-}{{$\leftarrow$}}1
        {bot}{{$\bot$}}1
        {LAMBDA}{{$\lambda$}}1
}

\lstdefinelanguage{scala}{
    morecomment = [s]{/*}{*/},
    morecomment = [l]{//},
    sensitive = true,
    morekeywords = {val,var,new,with,import,trait,this,def,if,else,Int},
    moredelim=**[is][\btHL]{§}{§},
}

\lstdefinelanguage{scheme}{
    sensitive = true,
    morekeywords={define,filter}
}

\lstdefinelanguage{sierra}{
    morecomment = [s]{/*}{*/},
    morecomment = [l]{//},
    morestring = [b]",
    sensitive = true,
    morekeywords = {uniform,varying,simd,scalar,for_each_active,for_each_unique,current_mask},
    morekeywords = {kernel,uint,mask,skip,true,false,uint32_t,uint64_t,nullptr,return,public,protected,private,template,auto,class,virtual,struct,union,void,this,size_t,volatile,if,else,do,while,case,goto,switch,for,while,bool,typedef,static,const,float,int,short,char,double,break,continue},
    keywords = {[2]define},
    keywordstyle={[2]\color{uds-purple}\bfseries},
    moredelim=**[is][\btHL]{§}{§},
}

\lstdefinelanguage{ssa}{
    sensitive = true,
    morekeywords={fn,bool,float,int,phi,goto,br,return},
    literate=
        {:=}{{$\gets$}}1
        {==}{{=}}1
        {!=}{{$\neq$}}1
        {<=}{{$\leq$}}1
        {>=}{{$\geq$}}1
        {->}{{$\rightarrow$}}1
        {<-}{{$\leftarrow$}}1
        {PHI}{{$\phi$}}1
}

\lstdefinelanguage{terra}{
    morecomment = [s]{/*}{*/},
    morecomment = [l]{//},
    sensitive = true,
    morekeywords = {int,function,if,then,return,else,elseif,terra,end,local},
    moredelim=**[is][\btHL]{§}{§},
}

\lstnewenvironment{lstinlinelisting}[1][]{%
  \noindent%
  \phantom{M}\minipage{\linewidth-.75em}%
  \lstset{frame=none,numbers=none,aboveskip=.5\bigskipamount,belowskip=.5\bigskipamount,backgroundcolor=,#1}%
}{%
  \endminipage%
}
\lstnewenvironment{lstintextlisting}[1][]{%
  \noindent%
  \phantom{M}\minipage{\linewidth-.75em}%
  \lstset{frame=none,numbers=none,basicstyle=\intextstyle,aboveskip=.5\bigskipamount,belowskip=.5\bigskipamount,backgroundcolor=,#1}%
}{%
  \endminipage%
}

\newcommand{\CopyrightNotice}{\hbox%
    {\parbox{18cm}{\textsf\centering\scriptsize$ $\\[-13cm]
    \centering
    This is a pre-print of an article accepted for publication in \emph{Proceedings of the International Conference on Field-Programmable Technology (FPT)}.\\
    \textcopyright~2021 IEEE.
    Personal use of this material is permitted.
    Permission from IEEE must be obtained for all other uses, in any current or future media, including reprinting/republishing this material for advertising or promotional purposes,creating new collective works, for resale or redistribution to servers or lists, or reuse of any copyrighted component of this work in other works.\par}%
    \vspace{-.85\baselineskip}}%
}

\begin{document}

\title{FLOWER: A Comprehensive Dataflow Compiler for High-Level Synthesis%
\thanks{This work is supported by the Federal Ministry of Education and Research (BMBF) as part of the HorME, HP-DLF, MetaDL, and REACT projects.}
}

\author{
    \IEEEauthorblockN{Puya Amiri\IEEEauthorrefmark{1},
    Arsène Pérard-Gayot\IEEEauthorrefmark{4},
    Richard Membarth\IEEEauthorrefmark{2}\IEEEauthorrefmark{1},
    Philipp Slusallek\IEEEauthorrefmark{1}\IEEEauthorrefmark{4},
    Roland Leißa\IEEEauthorrefmark{3},
    Sebastian Hack\IEEEauthorrefmark{4}}
    \IEEEauthorblockA{\IEEEauthorrefmark{1}German Research Center for Artificial Intelligence (DFKI), Germany}
    \IEEEauthorblockA{\IEEEauthorrefmark{2}Technische Hochschule Ingolstadt (THI), Research Institute AImotion Bavaria, Germany}
    \IEEEauthorblockA{\IEEEauthorrefmark{3}University of Mannheim (UMA), Germany}
    \IEEEauthorblockA{\IEEEauthorrefmark{4}Saarland University (UdS), Germany}
}
\maketitle

\CopyrightNotice{}

\begin{abstract}
\label{abstract}
FPGAs have found their way into data centers as accelerator cards, making reconfigurable computing more accessible for high-performance applications.
At the same time, new high-level synthesis compilers like Xilinx Vitis and runtime libraries such as XRT attract software programmers into the reconfigurable domain.
While software programmers are familiar with task-level and data-parallel programming, FPGAs often require different types of parallelism.
For example, data-driven parallelism is mandatory to obtain satisfactory hardware designs for pipelined dataflow architectures.
However, software programmers are often not acquainted with dataflow architectures---resulting in poor hardware designs.

In this work we present FLOWER, a comprehensive compiler infrastructure that provides automatic canonical transformations for high-level synthesis from a domain-specific library.
This allows programmers to focus on algorithm implementations rather than low-level optimizations for dataflow architectures.
We show that FLOWER allows to synthesize efficient implementations for high-performance streaming applications targeting System-on-Chip and FPGA accelerator cards, in the context of image processing and computer vision.
\end{abstract}


\section{Introduction}
Although Dennard scaling has broken down some time ago, it is generally assumed that Moore's law will continue to hold for at least a few years.
As a consequence, hardware vendors have built more and more specialized as well as parallel hardware such as multi-core CPUs, GPUs, or \mbox{FPGAs}.
Since FPGAs are low-power, reconfigurable and highly parallel integrated circuits, they have already been extensively adopted in embedded systems and more recently have found their way into scientific high-performance computing (HPC).

%High-performance computing (HPC) demands are gradually creating a technology gap between computational needs and performance.
%Even the abilities of modern multi- and many-core processors hardly bridge this gap.
%Additionally, post-Dennard scaling and the anticipated end of Moore’s law require new computing techniques.
%Although FPGAs have a significant share in embedded computing, only recently are they finding their place in scientific HPC beside GPUs.
Akin to languages for GPU computing such as CUDA or OpenCL, FPGA manufacturers offer various vendor-specific dialects of C/C++ that allow software developers to program at a high level of abstraction.
So-called \emph{high-level synthesis (HLS)} compiles these untimed, C-based dialects down into a timed, high-performance, register-transfer level (RTL) language in dataflow style.
These HLS languages entail two major drawbacks:
First, each dialect is closely tied to its vendor which makes code incompatible between different HLS languages.
Second, albeit HLS languages are typically C-based, they still require a hardware design mentality to be fully taken advantage of.
For example, transforming an untimed language into an RTL language calls for several transformations~\cite{9264692} with different levels of compilation and hardware synthesis flows.
These transformations have a different structure, depending on whether the input language is Xilinx C++ for HLS~\cite{9264692}, or Xilinx and Intel OpenCL~\cite{8891819}.

To sum up, FPGA vendors offer competing and incompatible HLS solutions.
For this reason, programmers must rewrite the application for each of those solutions.
Moreover, hardware and software interaction methods and optimizations are different among vendors and HLS dialects.

Most FPGA applications require some level of parallelism or concurrency to achieve the best performance, particularly for memory accesses.
HLS programming allows to express such parallelism through dataflow regions.
To take advantage of this, developers have to manually separate their application into tasks, and manually write all the necessary glue code to transform these separate tasks into a well-formed application.
This process is notoriously difficult, because it requires a lot of effort and knowledge of the application.

An alternative to HLS languages are domain-specific languages or libraries (DSLs):
By embedding the knowledge of a particular domain into a language, the compiler or library automatically applies efficient coding patterns, data movement mechanisms~\cite{8945776}, or spatial designs.

\paragraph*{Contributions}
This paper introduces FLOWER, a framework for FPGA development that makes the following contributions:
\begin{itemize}
    \item FLOWER provides a high-level syntax that helps in the design of dataflow-oriented FPGA applications---in particular by encouraging the separation between the core algorithm and data transfers.
    This simplifies low-level optimizations like vectorization or burst transfers.
    \item FLOWER automatically generates kernels that combine dataflow-oriented tasks from the dataflow graph of the application (see \cref{sec:top-level-gen}).
        While FLOWER's main target is HLS for Xilinx FPGAs, it can also generate multi-target OpenCL kernels which are optimized for both Intel and Xilinx FPGAs.
    \item FLOWER automatically generates host-code from a single piece of code that describes the entire application (see \cref{sec:host-code-gen}).
    \item We show the applicability of our framework on image processing and computer vision applications, where our framework has comparable or even better performance than alternatives (see \cref{sec:eval}).
\end{itemize}

\section{Background}
\subsection{Stages, Kernels and Tasks}
Typical streaming applications consist of several \emph{stages}.
For instance, \cref{applications} shows a list of applications with their respective number of stages. Considering a directed acyclic graph (DAG), each node of a DAG represents a stage.
In practice, using HLS, these stages are mapped to \emph{kernels} and \emph{tasks}.

A kernel is a function that is scheduled and controlled from the host code, and not from within the FPGA design.
On the contrary, a task is a function that is statically scheduled for execution from within the FPGA design, and not from the host. A kernel may contain one or more tasks.

Suppose we have a multi-stage application and its FPGA design consists of a single kernel: HLS tools will then apply a static model that will schedule every operation inside it.
After synthesis, different segments of the resulting hardware run in lockstep with each other, and cannot run concurrently.
While this coding style is simpler, it is not well-designed because dependencies or variable latency operations may introduce stalls.

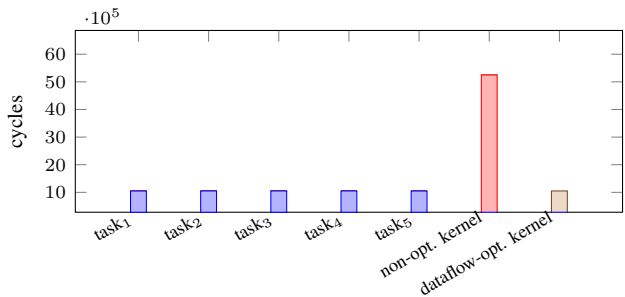
\begin{figure}
    \setlength{\abovecaptionskip}{0pt plus 0pt minus 0pt}
    \setlength{\belowcaptionskip}{0pt plus 0pt minus 0pt}
    \begin{tikzpicture}
    \begin{axis}
    [
        width=\columnwidth,
        height=4cm,
        ymin=1039000,
        ymax=6100000,
        scaled y ticks=base 10:-5,
        max space between ticks=10,
        ybar stacked,
        bar width = 6pt,
        enlargelimits = 0.15,
        ylabel={\footnotesize cycles},
        xtick={0,1,2,3,4,5,6,7,8},
        xticklabels={0,$\text{task}_1$,$\text{task}_2$,$\text{task}_3$,$\text{task}_4$,$\text{task}_5$,non-opt.\ kernel,dataflow-opt.\ kernel,8},
        xticklabel style={font=\scriptsize,rotate=30,anchor=east},
        yticklabel style={font=\scriptsize},
        ylabel near ticks,
        nodes near coords align={vertical},
    ]
        \addplot+ plot coordinates {
            (1, 1048576)
            (2, 1050632)
            (3, 1050632)
            (4, 1050632)
            (5, 1048577)
            (6, 0)
            (7, 0)
        };
        \addplot+ plot coordinates {
            (1, 0)
            (2, 0)
            (3, 0)
            (4, 0)
            (5, 0)
            (6, 5249058)
            (7, 0)
        };
        \addplot+ plot coordinates {
            (1, 0)
            (2, 0)
            (3, 0)
            (4, 0)
            (5, 0)
            (6, 0)
            (7, 1050644)
        };
    \end{axis}
    \end{tikzpicture}
    \caption{Latency in cycles for an FPGA design (frequency of 200 MHz) consisting of 5 tasks and one kernel. The two last bars shows the effect of dataflow optimization on a kernel containing these 5 tasks.}
    \label{fig:dataflow_motivation}
\end{figure}

A better approach is to apply a dataflow transformation that uses queues to transfer data between each task and enables task-parallelism.
With that approach, HLS tools will then generate a kernel that has a latency equal to the latency of the task with the highest latency.
This is in contrast to the previous approach, where the kernel had a latency equal to the sum of the latencies of each individual task.
\cref{fig:dataflow_motivation} shows the effect when applying dataflow optimization on a kernel that consists of five tasks.

The HLS compiler internally uses a Finite State Machine (FSM) to schedule individual parts of the kernel that does not use the dataflow transformation.
When an expensive operation is running, this FSM waits for its completion.
Hence, all other components of the kernel are in idle mode.
In cases where the kernel needs a significant amount of data, it may happen that the FPGA does not have enough BRAM to buffer them all, which means that the FPGA design may not function properly.
Moreover, such a kernel may need to access global memory with sporadic patterns, which may decrease the efficiency of the DMA engine.

In contrast to this, the dataflow-optimized kernel is made of several small tasks, which allows the HLS compiler to schedule each one independently, and generate one FSM per task.
This means that tasks have their own independent controllers, connected via FIFO buffers;
the buffering requirements get distributed among the tasks.
As a result, when a task stalls in a clock cycle, other tasks continue running as long as there is enough data in their input buffers, resulting in a higher overall throughput.
The dataflow transformation also has a significant impact on physical synthesis:
Shortening the critical paths allows the design to run at a higher clock frequency.
What is more, it benefits the fan-out of control signals.

\subsection{AnyHLS and FLOWER}
The work in this paper is built upon AnyHLS~\cite{oezkan2020anyhls}, a framework for FPGA application development that is itself built upon AnyDSL~\cite{leissa2018anydsl}.
AnyHLS introduces high-level abstractions to design FPGA applications, and extends the AnyDSL compiler infrastructure to generate FPGA designs for Intel OpenCL and Xilinx HLS.
For this, the syntax of AnyDSL is extended with additions for FPGA programming.
The image processing applications in AnyHLS are written in a library that builds on top of these changes.
This library allows programmers to develop point, local, and global image processing operators with very little effort.
In this paper, we focus on addressing the shortcomings of AnyHLS listed below, in particular with the automation of host code generation and dataflow optimizations.

AnyHLS provides a way to abstract typical patterns found in high-level synthesis in the form of a library with the help of the partial evaluator provided by AnyDSL.
These abstractions work well for single-kernel and single-task applications.
However, AnyHLS is limited to generate disjoint kernels in the form of IP-blocks without system integration, task-level pipelining, dataflow optimization, host-code generation, or memory optimizations such as burst transfers.
In fact, AnyHLS can only generate disjoint IPs from multi-stage applications, which then need manual wiring to connect them together to achieve a sequential execution.
In order to drive the design, the user has to write a corresponding host-side code for each application.
In FLOWER, we rely on AnyHLS abstractions to describe applications, and extend both AnyHLS and AnyDSL to support multi-stage applications by mapping them to different tasks and enable dataflow optimizations.
For this, FLOWER is deeply integrated into the AnyDSL compiler in order to apply task-level optimization and transformations.
We extend the AnyDSL compiler in order to extract the dataflow graph from application stages described by the user program and produce optimized dataflow pipelines according to the producer-consumer dependencies.
Unlike AnyHLS, our toolchain automates the whole design process from programming to synthesis.

\section{Motivation}
Software programming is very different from hardware design since traditional software design methods are not adapted to execution on FPGAs.
HLS tools have been introduced to help bridge that gap.
However, there are still areas where HLS does not help in the process of designing efficient hardware:

\subsection{Dataflow Transformations}
Vitis requires a canonical dataflow form to realize an architecture that takes advantage of task-pipelining and decreases redundant host-kernel communication.
This particular coding pattern is not only alien to software programmers, but it also demands a specific set of canonical rules which are difficult to apply, tedious, and may distract the programmer from implementing the actual algorithm.
This particularly applies for deep learning, image processing, and machine vision applications, where stages of a kernel need to be split into many tasks so that they can run concurrently or in parallel, and where a final kernel calls each task in the order dictated by the dataflow.
This is tedious and can be automated, since the order is in any case fixed by the dataflow.
%For instance, take the following HLS code:
%\begin{lstlisting}[language=c++]
%void read_gmem(int* in, hls::stream<int>& in_stream) {
%    for (int i=0; i<SIZE; ++i) in_stream << in[i];
%}
%
%void plus_one(
%    hls::stream<int>& in_stream,
%    hls::stream<int>& out_stream)
%{
%    for (int i=0; i<SIZE; ++i)
%        out_stream << (in_stream.read() + 1);
%}
%
%void write_gmem(int* out, hls::stream<int>& out_stream) {
%    for (int i=0; i<SIZE; ++i) out[i] = out_stream.read();
%}
%
%void increment(int* in, int* out) {
%    hls::stream<int> in_stream;
%    hls::stream<int> out_stream;
%#pragma HLS STREAM variable = in_stream depth = 10
%#pragma HLS STREAM variable = out_stream depth = 10
%#pragma HLS dataflow
%    read_gmem(in, in_stream);
%    plus_one(in_stream, out_stream);
%    write_gmem(out, out_stream);
%}
%\end{lstlisting}
%This code increments every element of an array, following the dataflow rules imposed by Vitis.
%This simple example splits the computation into 3 tasks that communicate via streams.
%Evidently, in order to use those tasks, the programmer has to combine all of them into a single kernel (called \lstinline{increment}).
%This, in turn, means writing annotations with \lstinline[language=c++]{#pragma}, and calling each individual task \emph{in the order} required by the data flowing between them.
%Doing this is highly error-prone.

\subsection{Low-level Optimizations}
\label{par:low-level-opts}
With FPGA hardware, unlike CPU architectures, the notion of a hierarchical memory that is transparent to the programmer does not exist.
Instead, there are plenty of BRAMs available on FPGA fabric that the HLS programmer must explicitly use to improve overall design performance and also increase global memory access efficiency~\cite{DBLP:journals/corr/abs-1807-01340}.
Considering that global memory access is taking a significant amount of time compared to kernel execution, missing a caching system becomes an essential problem.
In order to decrease this overhead and to exploit BRAMs as a cache, a batch process strategy must be used, typically with burst memory transfers.
Using burst memory transfers allows for minimizing the amount of control signal transactions and for merging several memory access requests into a single request.
This optimization greatly maximizes the application throughput and decreases global memory access latency.
Sadly, taking advantage of burst transfers requires to perform the dataflow transformation first, with all the above problems.

Another important low-level optimization for FPGA designs is \emph{vectorization}:
One goal of this optimization is to widen the bitwidth of the inputs of the kernel, so that many elements can be loaded at the same time and processed in parallel.
By this, vectorization increases throughput and memory efficiency.
%However, writing vectorized kernels is extremely difficult.
In order to vectorize the code, the HLS compiler requires consecutive memory access indices and several copies of the computation of interest.
Writing entire applications in this style is profoundly complex and error-prone.
In order to fully benefit from this optimization, the increase of the input's bitwidth should be matched by an appropriate number of units that process the data.
This will become substantially difficult if the application is not tailored for that.

\subsection{Interfacing Kernels with Hardware}
Hardware kernels and IPs need an interface to communicate with other hardware components and host devices via a handshaking protocol.
Depending on the parameters exposed by the hardware design to the outside or the way the design should be integrated into the application, HLS languages require various annotations and configurations.
%\input{listings/hls/HPC_interface}
%This kernel, for instance, expects 3 inputs: two arrays and one variable.
%The first array is connected via AXI (as shown by the keyword \lstinline{m_axi}) using the name \lstinline[language=c++]{gmem0}, while the second uses the name \lstinline[language=c++]{gmem1}.
%The last parameter \lstinline{var} is connected via AXI but uses the keyword \lstinline{s_axilite} since it is a scalar.
%The last \lstinline[language=c++]{#pragma} specifies the kernel execution mode (pipelined execution).
Writing the proper annotation for each kernel parameter makes the program fragile and bug-prone:
Any change to the parameters of that function has to be followed with an accompanying change to the corresponding annotations.

Another practical issue is writing the communication code between the host and the FPGA device.
Given a set of kernels, writing the corresponding host code in a separate file in order to interface them with the host is a strenuous endeavor on its own:
Programmers have to take care of buffer allocations, parameter types, and setting arguments, by using host-side APIs like XRT.
This requires a lot of boilerplate code, which is proportionately amenable for bugs and errors.

\section{FLOWER}
\label{sec:flower}
\begin{figure}
    \centering
    \includegraphics[width=0.45\textwidth]{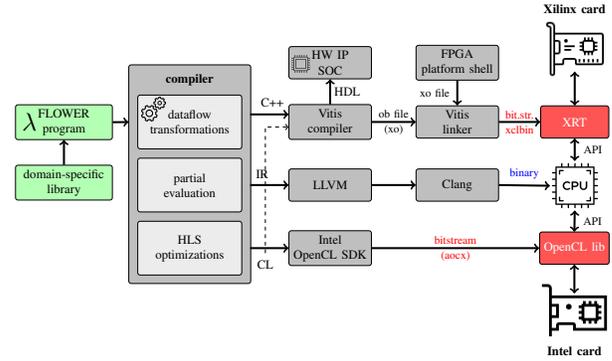}
    \caption{FLOWER: compilation workflow.}
    \label{fig:tool}
\end{figure}

The goal of FLOWER is to solve the problems discussed in the previous section.
\cref{fig:tool} shows the structure of FLOWER: There is only a single source program, from which both host and device code get generated.
Benefiting from Vitis features, this single source code can be simulated, emulated, or eventually synthesized to hardware.
%Configurable HLS compilation from software simulation to bitstream generation.

In order to explain the general workflow of FLOWER, let us have a look at a simple example:
\begin{lstlisting}
static mut chan1 : channel;
static mut chan2 : channel;
static mut chan3 : channel;
static mut chan4 : channel;

static vector_length = 4;

let in_img = read_image("input.png");
let (width, height) = (in_img.width, in_img.height);
let out_img = create_host_img(width, height);
let tmp_img1 = create_virtual_img(width, height, &mut chan3);
let tmp_img2 = create_virtual_img(width, height, &mut chan4);

let (in_img1, in_img2) = §split_image§(in_img, &mut chan1,
                                             &mut chan2);

for x, y, out, pix in §iteration_point§(tmp_img1, in_img1) {
    out.write(x, y, fun1(pix));
}

for x, y, out, pix in §iteration_point§(tmp_img2, in_img2) {
    out.write(x, y, fun2(pix));
}

for x, y, out, pix1, pix2 in
    §iteration_point2§(out_img, tmp_img1, tmp_img2) {
        out.write(x, y, fun3(pix1, pix2));
    }

image_write(out_img, "output.png");
\end{lstlisting}
This code uses the image processing DSL of AnyHLS~\cite{oezkan2020anyhls}.
DSL functions are highlighted in green.
Within FLOWER, each of these functions creates a task, resulting in 4 different tasks.
%one created by \lstinline{split_image}, two by \lstinline{iteration_point}, and one by \lstinline{iteration_point2}.
The function \lstinline{split_image} creates the first task, reads the input image \lstinline{in_img} from memory, and writes to two different virtual images (images that are mapped to channels).
Those virtual images are then used in two different point operators \lstinline{fun1} and \lstinline{fun2}, and the results are passed back to two more virtual images \lstinline{tmp_img1} and \lstinline{tmp_img2}.
Finally, a binary point operator \lstinline{fun3} is applied to the result of the two previous tasks, and the final result is written back to global memory.

The DSL functions all use the constant \lstinline{vector_length} internally, so that the resulting kernel is vectorized with burst transfers.
FLOWER achieves this by unrolling the computation within the loop body:
\begin{lstlisting}
fn @iteration_point(output: Img, input: Img,
                    body: fn(i32, i32, Img) -> ()) -> () {
    /* ... */
    for v in unroll(0, vector_length) {
        body(/* ... */);
    }
    /* ... */
}
\end{lstlisting}
This results in several copies of the for-loop body.
The HLS compiler is then able to determine the parts that can execute in parallel, resulting in the code being vectorized.

\subsection{Dataflow Graph Extraction}

From this example, FLOWER extracts a dataflow graph.
Each task represents a node in that graph and each channel is mapped to an edge.

FLOWER inspects each task to collect the channels that are read from (incoming edges) or written to (outgoing edges).
During this phase FLOWER detects invalid graphs and emits error messages if applicable.
In particular, it checks that the graph is acyclic and channels are written to or read from only once.
FLOWER generates the following graph from the example:

\begin{center}
    \vspace{-\baselineskip}
\begin{tikzpicture}[level distance=60mm,every node/.style={circle,inner sep=1pt, font = \small}]
    \node[shape=circle,draw=white,fill=blue!20,text=black] (split) at (0, 0) {$split$};
    \node[shape=circle,draw=white,fill=blue!20,text=black] (fun1) at (-1,-1) {$fun_1$};
    \node[shape=circle,draw=white,fill=blue!20,text=black] (fun2) at (1,-1)  {$fun_2$};
    \node[shape=circle,draw=white,fill=blue!20,text=black] (fun3) at (0,-2) {$fun_3$};

    \draw[-stealth] (0, 0.80) -- (split.north);
    \draw[-stealth] (fun3.south) -- (0, -2.80);

    \path [->](split) edge[bend right = 7] node[right]{} (fun1);
    \path [->](split) edge[bend left  = 7] node[left ]{} (fun2);
    \path [->](fun1)  edge[bend right = 7] node[right]{} (fun3);
    \path [->](fun2)  edge[bend left  = 7] node[right]{} (fun3);
\end{tikzpicture}
\end{center}

\subsection{Top-level Kernel Generation}
\label{sec:top-level-gen}

\begin{figure}
    \begin{subfigure}[t]{0.44\columnwidth}
        \resizebox{\textwidth}{!}{%
        \includegraphics[scale=1.03]{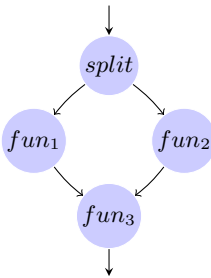}
        }
        \caption{before top-level generation}
        \label{subfig:cfg}
    \end{subfigure}%
    \begin{subfigure}[t]{0.58\columnwidth}
        \resizebox{\textwidth}{!}{%
        \includegraphics{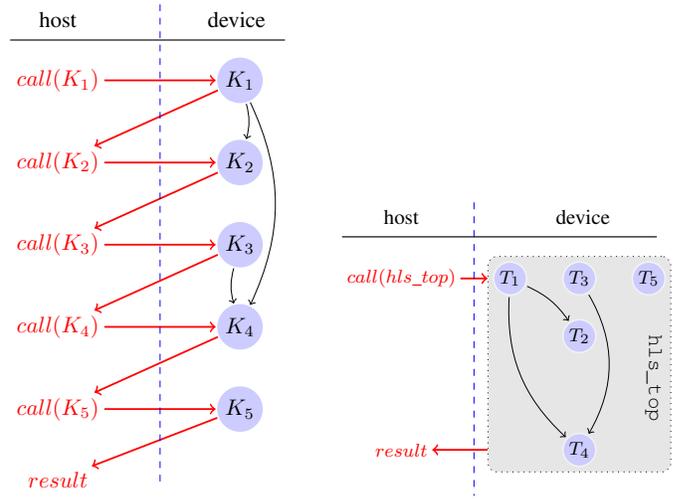}
        }
        \caption{after scheduling kernels as tasks}
        \label{subfig:cfg-opt}
    \end{subfigure}
    \caption{Control flow before top-level generation (a) and after scheduling kernels as tasks (b).}
    \label{fig:cfg}
\end{figure}

Our scheduling algorithm generates an HLS kernel that combines all the tasks of the application.
In the remainder of this text, we will refer to this kernel as the \emph{top-level kernel}.
For the HLS compiler to allow for concurrent or parallel execution of the tasks in the top-level kernel, FLOWER performs a topological sort of the graph in order to ensure that any task first writes to a channel before any tasks reads from that channel.
As a side note, this scheduling algorithm also works with tasks that are isolated from the rest of the graph.
Such tasks execute in parallel with the rest. \cref{fig:cfg} illustrates how the control flow changes between host/device code by introducing a top-level kernel.
On the device-side, FLOWER emits calls to each individual task and places appropriate \lstinline[language=c++]{#pragma} annotations such that the underlying HLS compiler picks up the dataflow region.
Here is, for instance, the generated top-level kernel for the example above:
\begin{lstlisting}[language=c++]
typedef struct { int e[4]; } int4;
typedef hls::stream<int4> int4_chan;

void task1(int[16], int4_chan*, int4_chan*) { /* ... */ }
void task2(int4_chan*, int4_chan*) { /* ... */ }
void task3(int4_chan*, int4_chan*) { /* ... */ }
void task4(int[16], int4_chan*, int4_chan*) { /* ... */ }

void hls_top(int input_data[16], int output_data[16]) {
#pragma HLS INTERFACE m_axi port = input_data
        bundle = gmem0 offset = slave
#pragma HLS INTERFACE s_axilite port = input_data
#pragma HLS STABLE variable = input_data
#pragma HLS INTERFACE m_axi port = output_data
        bundle = gmem0 offset = slave
#pragma HLS INTERFACE s_axilite port = output_data
#pragma HLS STABLE variable = output_data
#pragma HLS INTERFACE ap_ctrl_chain port = return
#pragma HLS top name = hls_top
#pragma HLS DATAFLOW

    int4_chan chan1_slot, chan2_slot, chan3_slot, chan4_slot;
    int4_chan* chan1 = &chan1_slot;
    int4_chan* chan2 = &chan2_slot;
    int4_chan* chan3 = &chan3_slot;
    int4_chan* chan4 = &chan4_slot;
#pragma HLS STREAM variable = chan1 depth = 2
#pragma HLS STREAM variable = chan2 depth = 2
#pragma HLS STREAM variable = chan3 depth = 2
#pragma HLS STREAM variable = chan4 depth = 2
    task1(input_data, chan1, chan2);
    task2(chan1, chan3);
    task3(chan2, chan4);
    task4(output_data, chan3, chan4);
}
\end{lstlisting}
Note that this code uses channels of type \lstinline{int4}, since we have a vectorization factor of 4.
FLOWER generates separate tasks as separate functions.
The tasks \lstinline{task1} and \lstinline{task4} have a parameter that allows them to access global memory.
For the same reason the top-level kernel \lstinline{hls_top} expects two parameters: \lstinline{input_data} and \lstinline{output_data}.
FLOWER annotates these parameters with pragmas to instruct the underlying HLS compiler to give them an AXI interface to connect to other peripherals.
FLOWER defines the 4~channels \lstinline{chan1} to \lstinline{chan4} as FIFO channels to communicate data between tasks (using the \lstinline[language=c++]{#pragma HLS STREAM} annotation).
Finally, we see how FLOWER places calls of these tasks in topological order as discussed previously and tells the HLS compiler via \lstinline[language=c++]{#pragma HLS DATAFLOW} of a dataflow region.
Consequently, this structure results in a design in which all tasks are pipelined and execute concurrently.

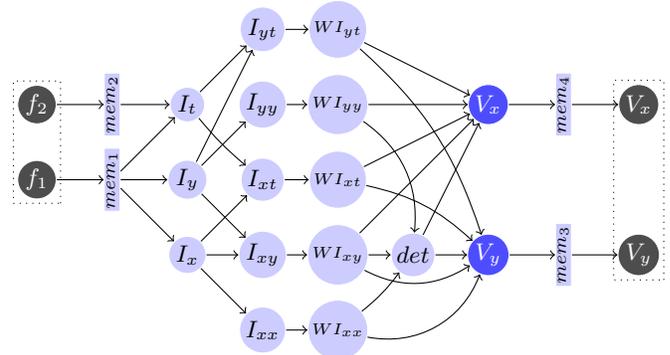
\begin{figure}
    \begin{minipage}{0.48\textwidth}
        \begin{turn}{90}
            \centering
            \begin{tikzpicture}
%grow = down,
%level distance = 200mm,
%sibling distrnce = 200mm]
[grow = down, level distance = 60mm,
every node/.style={circle,inner sep=1pt, font = \small, rotate=270}
%level 1/.style = {sibling distance = 20mm,nodes = {fill = red!45}},
%level 2/.style = {sibling distance=10mm,nodes={fill = red!30}},
%level 3/.style = {sibling distance = 5mm,nodes = {fill = red!25}}
%level 3/.style = {sibling distance = 900mm}
]
    \node[shape = circle,draw = white, fill = black!70, text=white ] (A) at (0, 0) {$f_1$};
    \node[shape = rectangle,draw = white, fill = blue!20, text=black,font=\fontsize{7}{5},rotate=90 ] (M1) at (0, -1) {$mem_1$};
    \node[shape = circle,draw = white, fill = black!70, text=white ] (B) at (1,0)  {$f_2$};
    \node[shape = rectangle,draw = white, fill = blue!20, text=black,font=\fontsize{7}{5},rotate=90 ] (M2) at (1, -1) {$mem_2$};
    \node[shape = circle,draw = white, fill = blue!20, text=black ] (C) at (-1,-2) {$I_{x}$};
    \node[shape = circle,draw = white, fill = blue!20, text=black ] (D) at (0,-2)  {$I_{y}$};
    \node[shape = circle,draw = white, fill = blue!20, text=black ] (E) at (1,-2)  {$I_{t}$};
    \node[shape = circle,draw = white, fill = blue!20, text=black ] (F) at (-2,-3) {$I_{xx}$};
    \node[shape = circle,draw = white, fill = blue!20, text=black ] (G) at (0,-3)  {$I_{xt}$};
    \node[shape = circle,draw = white, fill = blue!20, text=black ] (H) at (-1,-3) {$I_{xy}$};
    \node[shape = circle,draw = white, fill = blue!20, text=black ] (I) at (1,-3)  {$I_{yy}$};
    \node[shape = circle,draw = white, fill = blue!20, text=black ] (J) at (2,-3)  {$I_{yt}$};
    \node[shape = circle,draw = white, fill = blue!20, text=black, font=\fontsize{6}{0}] (R) at (-2,-4) {$WI_{xx}$};
    \node[shape = circle,draw = white, fill = blue!20, text=black, font=\fontsize{6}{0} ] (K) at (-1,-4) {$WI_{xy}$};
    \node[shape = circle,draw = white, fill = blue!20, text=black, font=\fontsize{6}{0} ] (L) at (0,-4) {$WI_{xt}$};
    \node[shape = circle,draw = white, fill = blue!20, text=black, font=\fontsize{6}{0} ] (M) at (1,-4) {$WI_{yy}$};
    \node[shape = circle,draw = white, fill = blue!20, text=black, font=\fontsize{6}{0} ] (N) at (2,-4) {$WI_{yt}$};
    \node[shape = circle,draw = white, fill = blue!20, text=black ] (O) at (-1,-5) {$det$};
    \node[shape = circle,draw = white, fill = blue!70, text=white, font=\fontsize{9}{0}] (P) at (1,-6) {$V_{x}$};
    \node[shape = circle,draw = white, fill = blue!70, text=white, font=\fontsize{9}{0}] (Q) at (-1,-6) {$V_{y}$};
    \node[shape = rectangle,draw = white, fill = blue!20, text=black,font=\fontsize{7}{5},rotate=90 ] (M3) at (-1, -7) {$mem_3$};
    \node[shape = rectangle,draw = white, fill = blue!20, text=black,font=\fontsize{7}{5},rotate=90 ] (M4) at (1, -7) {$mem_4$};
    \node[shape = circle,draw = white, fill = black!70, text=white, font=\fontsize{9}{0}] (O1) at (1,-8) {$V_{x}$};
    \node[shape = circle,draw = white, fill = black!70, text=white, font=\fontsize{9}{0}] (O2) at (-1,-8) {$V_{y}$};
    \node[draw, dotted, inner sep=0.5mm, shape=rectangle, fit=(A) (B),rotate=90] {};
    \node[draw, dotted, inner sep=0.5mm, shape=rectangle, fit=(O1) (O2),rotate=90] {};

    \path [->](A) edge node[right] {} (M1);
    \path [->](M1) edge node[right] {} (D);
    \path [->](M1) edge node[left]  {} (C);
    \path [->](M1) edge node[left]  {} (E);
    \path [->](B) edge node[right] {} (M2);
    \path [->](M2) edge node[right] {} (E);
    \path [->](C) edge node[right] {} (F); 
    \path [->](C) edge node[right] {} (G); 
    \path [->](C) edge node[right] {} (H); 
    \path [->](D) edge node[right] {} (I); 
    \path [->](D) edge node[right] {} (H); 
    \path [->](D) edge node[right] {} (J); 
    \path [->](E) edge [bend right = 5] node {} (G); 
    \path [->](E) edge [bend right = 0] node {} (J); 
    \path [->](H) edge node [right]{} (K); 
    \path [->](R) edge [bend right = 7] node {} (O); 
    \path [->](K) edge node [right]{} (O); 
    \path [->](M) edge [bend left=30] node {} (O); 
    \path [->](F) edge [right] node {} (R); 
    \path [->](G) edge node [right]{} (L); 
    \path [->](I) edge node [right]{} (M); 
    \path [->](J) edge node [right]{} (N); 
    \path [->](R) edge [bend right = 40] node {} (Q); 
    \path [->](K) edge node [right]{} (P); 
    \path [->](K) edge [bend right = 30] node {} (Q); 
    \path [->](L) edge node [right]{} (P); 
    \path [->](L) edge [bend left = 15] node {} (Q); 
    \path [->](M) edge node [right]{} (P); 
    \path [->](N) edge node [right]{} (P); 
    \path [->](N) edge [bend left =15] node {} (Q); 
    \path [->](O) edge node [right]{} (Q); 
    \path [->](O) edge node [right]{} (P); 
    \path [->](Q) edge node [right]{} (M3); 
    \path [->](P) edge node [right]{} (M4); 
    \path [->](M3) edge node [right]{} (O2); 
    \path [->](M4) edge node [right]{} (O1); 

\end{tikzpicture}
        \end{turn}
        \caption{Data flow graph for the Lucas-Kanade implementation for optical flow estimation. $f_1$ and $f_2$ denote two unique frames. $V_x$ and $V_y$ are components of motion vectors. $I_x$ and $I_y$ are spatial derivatives. $I_t$ is a temporal derivative. $WI_{xy}$ is an example of windowed weighted averages. Splitting nodes are removed for the sake of simplicity. $mem_{1-4}$ represent 4 different memory interfaces.}
        \label{fig:opticalflow_lk}
    \end{minipage}
\end{figure}

While this example plainly introduces the fundamental functionality of our toolchain, FLOWER is not limited to that.
\Cref{fig:opticalflow_lk} introduces another example.
It demonstrates a more complicated dataflow graph that implements the Lucas-Kanade method for optical flow estimation.
Black nodes are not part of algorithm, they specify inputs and outputs and reside on the host-side.
Splitting nodes are not shown for the sake of simplicity.
Since there are parallel paths from inputs to outputs, a single memory interface cannot feed the tasks concurrently.
Squared nodes named $mem_{1-4}$ solve this issue.
FLOWER designates 4 different memory bundles using interface pragmas to separate memory transactions. Assigning individual memory interfaces avoid congestion in host-device memory transfers.

\subsection{Hardware/Software Interface}
\label{sec:host-code-gen}

In order to use the generated FPGA design in a practical setting, we need to interface it with a host.
FLOWER generates interface pragmas for different target platforms.
However, this is not sufficient because the HLS code on its own does not specify how to communicate data from or to the host system.
In order to do that, the HLS code has to be driven by a host code, that is typically written with the XRT API provided by Xilinx.
Our framework generates such host code automatically.
The generated code contains the necessary XRT API calls required to launch the kernel and communicate with it.
For instance, for the application above, the following equivalent host code will be automatically generated (our framework generates the host code as LLVM IR, not C++, but the concepts are the same):
\begin{lstlisting}[language=c++]
auto device = xcl::get_devices()[0];
auto bitsteam_buffer =
    xcl::read_binary_file("fpga_bitsream.xclbin");
cl::Program::Binaries bins
    {{ bitstream_buffer.data(), bitstream_buffer.size() }};
auto context = cl::Context(device, NULL, NULL, NULL);
auto q = cl::CommandQueue(context, device, 0);

cl::Buffer buffer_input(context, CL_MEM_READ_WRITE);
cl::Buffer buffer_output(context, CL_MEM_READ_WRITE);

auto [input_data, width, height] = load_png("input.png");
q.enqueueWriteBuffer(buffer_input, true,
                     0, width * height, input_data);

cl::Program program(context, {device}, bins, NULL);
auto kernel = cl::Kernel(program, "hls_top");
kernel.setArg(0, buffer_input);
kernel.setArg(1, buffer_output);
q.enqueueTask(kernel);
q.finish();

auto output_data = alloc_pixels(width, height);
q.enqueueReadBuffer(buffer_output, true,
                    0, width * height, output_data);
write_png(output_data, width, height, "output.png");
\end{lstlisting}
%For instance, for the application above, FLOWER will generate host code that sets up the basic infrastructure to load the kernel, creates the OpenCL/XRT context and command queue, then creates buffers to hold the input data and the output data, loads the image, runs the kernel, and finally writes back the result image.

Concisely, this code sets up the basic infrastructure to load the kernel, creates the OpenCL/XRT context and command queue, then creates buffers to hold the input and output data, loads the image, runs the kernel, and finally writes back the result image.
In order to generate that code, FLOWER considers every loop that comes from the DSL (for instance, the loops created via \lstinline{iteration_point}) as executing on the FPGA.
Thus, these parts are translated into a single launch of the top-level kernel.
The rest is considered as running \emph{on the host}: The calls to \lstinline{read_image} or \lstinline{write_image}, for instance, will be executed there, and not on the FPGA.
Internally, functions like \lstinline{write_image} or \lstinline{read_image} use compiler-provided intrinsics to copy data from the host to the FPGA:
Those directly translate to calls to XRT that transfer the data in the right direction.

As mentioned previously, all the loops that are generated via the DSL translate into one top-level kernel launch on the host.
The arguments of that kernel launch are set according to the parameters extracted during the top-level kernel generation phase.
Those typically come from uses of input or output images in the DSL loops, like in this example.

Thanks to that automatic host code generation, the programmer only needs to focus on writing the application from a single piece of code written using FLOWER.
Consequently, it's easier to make modifications of the code, since the host code is automatically synchronized with the FPGA code.

% write about the effect of each transformations, data-parallelism and task parallelism, interface chaining, dataflow from host to device
% All examples in fig4 except AnyHLS(base) are dataflow optimized. mention it
% we removed all technical details while reviewers asked for more technical details.
% Put a more complicated graph
% adding our AnyHLS vs FLOWER description in introduction?

\section{Evaluation}
\label{sec:eval}
For experimental evaluation, we consider a range of prominent applications that have been used in comparable works~\cite{10.1145/3373087.3375320,7756767,10.1145/3377555.3377899,10.1145/2897824.2925892}.
\cref{applications} lists the number of stages of each application.
This number does not include two additional memory read/write stages required for burst transfer optimization.

\begin{table}[tbh!]
    \caption{Benchmarking applications.}
    \label{applications}
    %\resizebox{\columnwidth}{!}{%
    {\scriptsize
    \begin{tabular}{lcl}
        \toprule
        application& stage(s) & description\\
        \midrule
        Mean filter & 1 & $5\times 5$ filter reducing intensity variation\\
        Gaussion blur & 1 & $5\times 5$ integer low-pass filter for noise reduction\\
        Bilateral filter & 1 & $5\times 5$ floating-point filter for image smoothing\\
                         &   & while preserving edges\\
        Sobel-Luma & 2 & Edge detection algorithm utilizing\\
                   &   & RGB to luma color conversion\\
        Unsharp mask & 3 & Sharpens an image\\
        Filter chain & 3 & $3 \times 3$ filter chained 3 times\\
        Jacobi & 1 & $3\times 3$ filter for image smoothing\\
        Optical flow (LK) & 16 & Lucas-Kanade method for motion estimation\\
        %Optical flow (HS) & 30 & Horn Schunck iterative method for motion estimation\\
        Harris & 9 & Corner detection for finding features in images\\
        Shi-Thomasi & 9 & Corner detection with improved scoring function\\
        Laplace & 1 & Derivative operator for edge detection\\
        Square & 1 & Pixel-wise operation for increasing image contrast\\
        Sobel & 1 & $3\times 3$ filter for edge detection\\
        \bottomrule
    \end{tabular}
    }
    %}
\end{table}

We evaluate FLOWER on two different FPGA platforms:
Xilinx Alveo U280 (xcu280-fsvh2892-2L-e) and Bittware 520N-MX (Intel Stratix 10 MX2100).
Both are accelerator cards connected to the host via PCIe Gen3x16.
However, we can only use PCIe Gen3x8 for the Bittware 520N-MX due to restrictions of the used BSP.

Intel OpenCL codes are synthesized by the Intel FPGA SDK for OpenCL 19.4.
Xilinx HLS C++ and OpenCL codes for the accelerator card are synthesized by the Vitis v++ compiler 2020.1.
We use Vitis\_hls 2020.1 to synthesize the generated IPs.
Host programs for the Xilinx card use the XRT 2.7.766 runtime library.
\cref{benchmark_index} shows available benchmark evaluations with corresponding plots among different backends of FLOWER.
%\begin{table}[tbh!]
\begin{table}
    %\documentclass[varwidth]{standalone}
%\usepackage[backend=biber]{biblatex}
%\bibliography{references.bib}
%\usepackage{amsmath}
%\usepackage{cleveref}
%\usepackage{pifont}
%\begin{document}

\begin{tabular}{l*{4}{c}}
\toprule
\multicolumn{5}{c}{Xilinx}\\
\cline{2-4}
              & HLS & HLS-SoC & OpenCL & Intel OpenCL\\
\midrule
Mean filter      &\ding{51}&\ding{51}&    -     &\ding{51} \\
Gaussian blur    &\ding{51}&\ding{51}& \ding{51}&\ding{51} \\
Bilateral filter &\ding{51}&\ding{51}&    -     &   -      \\
Sobel-Luma       &\ding{51}&\ding{51}&    -     &   -      \\
Unsharp mask     &\ding{51}&   -     &    -     &   -      \\
Filter chain     &\ding{51}&   -     &    -     &\ding{51} \\
Jacobi           &\ding{51}&   -     &    -     &\ding{51} \\
Optical flow (LK)&\ding{51}&   -     &    -     &   -      \\
Harris           &\ding{51}&\ding{51}&    -     &\ding{51} \\
Shi-Thomasi      &\ding{51}&   -     &    -     &   -      \\
Laplace          &\ding{51}&   -     &    -     &   -      \\
Square           &\ding{51}&   -     &    -     &   -      \\
Sobel            &\ding{51}&\ding{51}&    -     &   -      \\
\midrule
Figure number    &\cref{plot:xil_opt_hls}&\cref{plot:hipacc_vs_flower}&\cref{plot:xil_ocl_gaussian}& \cref{plot:intel_ocl}\\
\bottomrule
\end{tabular}

%\end{document}

    \caption{Available application evaluations for the different backends in FLOWER.}
    \label{benchmark_index}
\end{table}

% GOALS:
% - FLOWER IS COMPARABLE/FASTER THAN ALTERNATIVES (PERFORMANCE)
% - FLOWER IS EASIER TO USE THAN ALTERNATIVES (PRODUCTIVITY ADVANTAGES)
% - FLOWER RUNS ON INTEL (SAME CODE, MANY PLATFORMS)
We start by evaluating our framework against Hipacc~\cite{membarth2016hipacc,reiche2017hipaccfpga} and AnyHLS~\cite{oezkan2020anyhls} on a set of image processing applications, before assessing the OpenCL support.

\subsection{Hipacc}% vs. Hipacc
\begin{figure}
    \centering
    \begin{tikzpicture}
    \begin{axis}
    [
        small,
        %ymajorgrids = true,
        width=0.70\columnwidth,
        height=4.3cm,
        %width=\textwidth,
        %height=.5\textwidth,
        ymin=1047000,
        ymax=1057000,
        scaled y ticks=base 10:-5,
        max space between ticks=9,
        ybar = 1pt,
        bar width = 6pt,
        enlargelimits = 0.15,
        ylabel={\footnotesize cycles},
        ylabel near ticks,
        xlabel={},
        symbolic x coords={
            gaussian,
            %gaussian\_v4,
            laplace,
            mean filter,
            %laplace\_v4,
            sobel,
            %sobel\_4,
            harris corner,
            %Harris corner\_v4,
            bilateral,
            %bilateral\_v4
        }, % these are the specification of coordinates on the x-axis.
        xtick=data,
        %nodes near coords, % this command is used to mention the y-axis points on the top of the particular bar.
        nodes near coords align={vertical},
        xticklabel style={font=\footnotesize,rotate=30,anchor=base,xshift=-0.50cm,yshift=-0.23cm},
        yticklabel style={font=\footnotesize,rotate=00,anchor=base,xshift=-0.35cm,yshift=0cm},
        %legend pos=north east,
        %legend style={cells={align=left,anchor=west},font=\scriptsize},
        %legend style={at={(0.5,0.9)},anchor=north,legend columns=-1,font=\scriptsize},
        legend style={anchor=north east,draw=none,fill=none,at={(current axis.north east)},legend columns=-1,font=\scriptsize},
    ]
    %FLOWER
    \addplot coordinates {
    (gaussian,         1050634)
    (laplace,           1049607)
    (mean filter,       1050636)
    (sobel,             1049610)
    (harris corner,     1049622)
    (bilateral,         1050875)
    };
    %Hipacc
    \addplot coordinates {
    (gaussian,          1052686)
    (laplace,           1050632)
    (mean filter,       1052688)
    (sobel,             1050632)
    (harris corner,     1050642)
    (bilateral,         1052930)
    };
    \addlegendentry{FLOWER}
    \addlegendentry{Hipacc}
    \end{axis}
    \end{tikzpicture}%
    \begin{tikzpicture}
    \begin{axis}
    [
        small,
        width=3.5cm,
        height=4.3cm,
        ymin=262000,
        ymax=265000,
        scaled y ticks=base 10:-5,
        max space between ticks=10,
        ybar = 1pt,
        bar width = 6pt,
        enlarge x limits = 0.3,
        xlabel={},
        symbolic x coords={
            laplace,
            sobel,
            harris corner,
        }, % these are the specification of coordinates on the x-axis.
        xtick=data,
         %nodes near coords, % this command is used to mention the y-axis points on the top of the particular bar.
        nodes near coords align={vertical},
        xticklabel style={font=\footnotesize,rotate=30,anchor=base,xshift=-0.50cm,yshift=-0.23cm},
        yticklabel style={font=\footnotesize,rotate=00,anchor=base,xshift=-0.25cm,yshift=0cm},
    ]
    %FLOWER
    \addplot coordinates {
    (laplace,       262405)
    (sobel,          262405)
    (harris corner, 262561)
    };
    %Hipacc
    \addplot coordinates {
    (laplace,       263432)
    (sobel,          263432)
    (harris corner, 263170)
    };
    \end{axis}
    \end{tikzpicture}
    \caption{Synthesis results, showing the latency of applications generated by FLOWER and Hipacc. Image size is $1024 \times 1024$. $f_\mathit{target}=300\mathit{MHz}$ on a Xilinx Alveo U280 card. Left shows non-vectorized version, right is vectorized with a factor of 4.}
    \label{plot:hipacc_vs_flower}
   \end{figure}
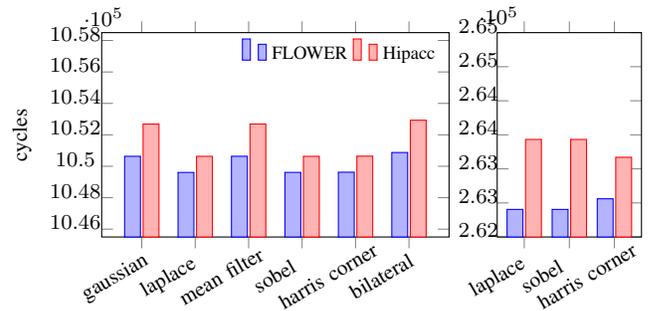

The FPGA support in Hipacc is mostly designed for Zynq SoPC (System on Programmable Chip) platforms, and the generated IP blocks obtained from Hipacc are not immediately ready to be linked with the accelerators' platform shell.
Therefore, to compare our work with Hipacc, we rely on the SoPC IP output of FLOWER that is synthesized for the FPGA part \emph{xcu280-fsvh2892-2L-e} found in the Alveo U280 card.
With Hipacc, we generate each application by first making it compatible with the \verb$vitis_hls$ command-line tool, and then by synthesizing it as streaming IPs for the same FPGA part.
All applications in Hipacc have an AXIS interface without any global memory control bundle, and thus they cannot access global memory.
Consequently, we need to impose this restriction on FLOWER applications as well.
Synthesis results in \cref{plot:hipacc_vs_flower} alongside with resource usage in \cref{table:hipacc_vs_flower_resource} shows that FLOWER applications have lower latencies. This is true also when applications are vectorized.

\begin{table}[ht]
\centering
\setlength\tabcolsep{4pt}
\caption{Post place and route resource usage of FLOWER and Hipacc applications. Image size is $1024 \times 1024$. Reported by Vitis targeting Xilinx Alveo U280 card.}
\label{table:hipacc_vs_flower_resource}
\begin{tabular}[t]{llrrrrrr}
\toprule
Application & Tool & CLB & LUT & FF & DSP & BRAM & SRL\\
\midrule
\multirow{2}{*}{Gaussian} & FLOWER & 363 & 1851 & 1486 & 0 & 8 & 32\\
&Hipacc &           602 & 2987 & 2503 & 0 & 8 & 0\\
\midrule
\multirow{2}{*}{Laplace} & FLOWER & 99 & 371 & 487 & 0 & 4 & 32\\
&Hipacc &          86 & 374 & 454 & 0 & 4 & 0\\
\midrule
\multirow{2}{*}{Mean filter} & FLOWER & 396 & 1861 & 1509 & 4 & 8 & 32\\
&Hipacc &          634 & 3257 & 2413 & 4 & 8 & 0\\
\midrule
\multirow{2}{*}{Sobel} & FLOWER & 258 & 1124 & 1156 & 0 & 8 & 72\\
&Hipacc &        329 & 1410 & 1721 & 0 & 8 & 0\\
\midrule
\multirow{2}{*}{Harris corner} & FLOWER & 712 & 3337 & 3656 & 22 & 20 & 240\\
&Hipacc &               1298 & 6241 & 7098 & 21 & 20 & 217\\
\midrule
\multirow{2}{*}{Bilateral} & FLOWER & 18189 & 76708 & 81909 & 1097 & 8 & 8062\\
&Hipacc &           11781 & 51644 & 57906 & 654 & 8 & 5960\\
\bottomrule
\end{tabular}
\end{table}

\subsection{AnyHLS}% vs. AnyHLS
AnyHLS does not use any dataflow optimization.
That is, for applications with multiple stages like filter-chain or Harris-corner, AnyHLS only generates opaque modules.
These modules cannot communicate to each other or to the host.
Therefore, a huge amount of manual HLS coding is required to optimize and connect them together properly.
Applications generated by FLOWER work without any manual optimization and produce superior results.

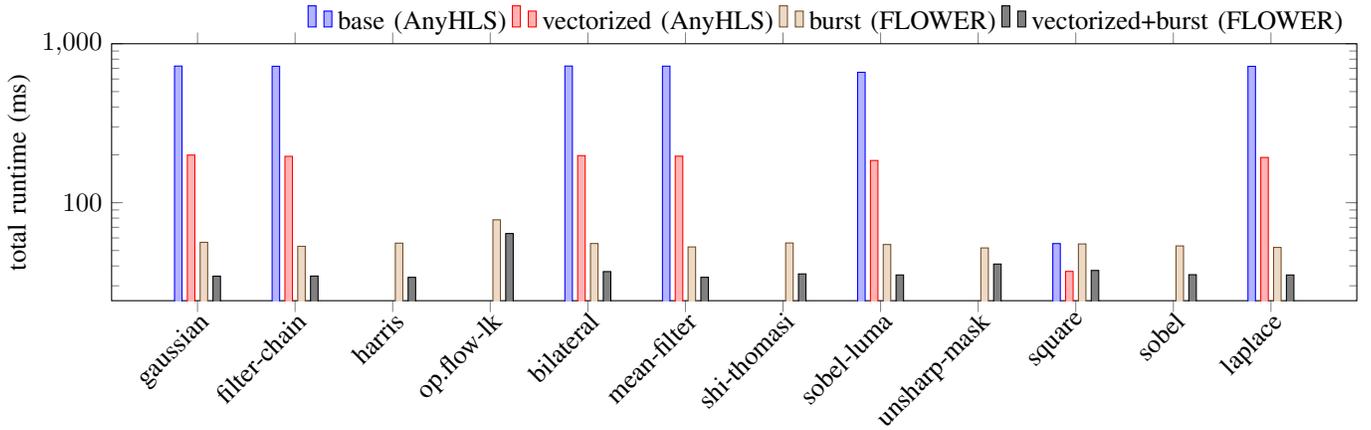
\begin{figure*}
\setlength{\abovecaptionskip}{0pt plus 0pt minus 0pt}
\setlength{\belowcaptionskip}{0pt plus 0pt minus 0pt}
\begin{tikzpicture}
    \begin{axis}[
        width = \textwidth, height = 5cm,
        ybar,
        ymax=1000,
        %ymin=0,
        %scaled y ticks=base 10:-2,
        ymode=log,
        log basis y={10},
        log ticks with fixed point,
        %max space between ticks=50,
        legend columns = -1, % Horizontal legend
        %legend style = { at = {(0.5, -0.51)}, anchor = north}, % Place legend below the graph
        legend style={anchor=south east,draw=none,fill=none,at={(current axis.north east)}},
        %symbolic x coords = { filter-chain, harris, op.flow-lk, bilateral, mean-filter, shi-thomasi, op.flow-hs, sobel-luma, unsharp-mask, square, sobel, laplace },
        symbolic x coords = {gaussian, filter-chain, harris, op.flow-lk, bilateral, mean-filter, shi-thomasi, sobel-luma, unsharp-mask, square, sobel, laplace },
        xtick = {gaussian, filter-chain, harris, op.flow-lk, bilateral, mean-filter, shi-thomasi, sobel-luma, unsharp-mask, square, sobel, laplace},
        x tick label style={yshift=0.20cm,rotate=45,anchor=north east}, % Rotate horizontal ticks by 45 degrees to save space
        enlarge x limits = 0.08,
        ylabel = {total runtime (ms)},
        bar width = 1mm,
        ]
        \addplot coordinates {(gaussian, 723.7)   (filter-chain, 721.728) (bilateral, 724.02)  (mean-filter, 723.001) (sobel-luma, 661.441) (square, 55.308) (laplace, 720.962)}; %
        \addplot coordinates {(gaussian, 199.424) (filter-chain, 195.702) (bilateral, 197.755) (mean-filter, 196.226) (sobel-luma, 184.271) (square, 36.96)  (laplace, 192.617)}; % _v
        %\addplot coordinates { (filter-chain, 53.136) (harris, 55.655) (op.flow-lk, 78.004) (bilateral, 55.353) (mean-filter, 52.676) (shi-thomasi, 55.791) (op.flow-hs, 79.942) (sobel-luma, 54.563) (unsharp-mask, 51.919) (square, 55.01) (sobel, 53.464) (laplace, 52.32) }; % _burst
        \addplot coordinates {(gaussian, 56.3) (filter-chain, 53.136) (harris, 55.655) (op.flow-lk, 78.004) (bilateral, 55.353) (mean-filter, 52.676) (shi-thomasi, 55.791) (sobel-luma, 54.563) (unsharp-mask, 51.919) (square, 55.01) (sobel, 53.464) (laplace, 52.32) }; % _burst
        %\addplot coordinates { (filter-chain, 53.136) (harris, 55.655) (op.flow-lk, 78.004) (bilateral, 55.353) (mean-filter, 52.676) (shi-thomasi, 55.791) (sobel-luma, 54.563) (unsharp-mask, 51.919) (square, 55.01) (sobel, 53.464) (laplace, 52.32) }; % _burst
        \addplot coordinates {(gaussian, 34.5) (filter-chain, 34.559) (harris, 33.915) (op.flow-lk, 63.983) (bilateral, 36.852) (mean-filter, 33.939) (shi-thomasi, 35.642) (sobel-luma, 35.102) (unsharp-mask, 41.113) (square, 37.473) (sobel, 35.242) (laplace, 35.102) }; % _burst_v
        %\addplot coordinates { (filter-chain, 13.385) (harris, 13.02) (bilateral, 14.336) (mean-filter, 13.282) (shi-thomasi, 12.912) (sobel-luma, 12.981) (unsharp-mask, 19.887) (square, 12.0981) (sobel, 14.069) (laplace, 13.652) }; % _burst_v_host
        \legend{base (AnyHLS), vectorized (AnyHLS), burst (FLOWER), vectorized+burst (FLOWER),\begin{comment} vectorized+burst+enhanced\end{comment} host}
    \end{axis}
\end{tikzpicture}
\caption{Total kernel runtime of applications through different optimizations, when executed 6 times on a Xilinx Alveo U280. Image size is $1024\times 1024$ and vectorization factor is 4. The AnyHLS versions have no burst transfer. Applications which do not have an AnyHLS version could not be generated using the AnyHLS Xilinx backend, because for the generated kernels, the synthesis tool requires FIFO buffer sizes that are too large.}
\label{plot:xil_opt_hls}
\end{figure*}

\begin{figure}
    \centering
    %\documentclass{standalone}
%\usepackage{tikz,pgf}
%\usetikzlibrary{fit,graphs,backgrounds,scopes}

\def\distancelabel{0.01}
\tikzstyle{textbox} = [rectangle,draw=none,font=\small,align=center]%
\tikzstyle{codebox} = [rectangle,draw=none,font=\ttfamily\small,align=center]%
\tikzstyle{coord} = [coordinate,node distance=\distancex,align=center]%
\tikzstyle{outerbox} = [rectangle,draw=black,dashed,rounded corners]%

%\begin{document}
    \begin{tikzpicture}
        \node[shape=circle,draw=white,fill=red!70, text=white] (A) {$T_{R}$};
        \node[shape=circle,draw=white,fill=blue!20,text=black] (B) [right=of A] {$T_{1}$};
        \node[shape=circle,draw=white,fill=blue!20,text=black] (C) [right=of B] {$T_{2}$};
        \node[shape=circle,draw=white,fill=red!70, text=white] (E) [right=of C] {$T_{W}$};

        \path[->,thick](A) edge node[left]  {} (B);
        \path[->,thick](B) edge node[right] {} (C);
        \path[->,thick](C) edge node[right] {} (E);

        \begin{scope}[on background layer]
            \node[outerbox,dotted,fill=gray!50,inner sep=0.5mm,shape=rectangle] (GRAY) [fit=(B) (C)] {};
            \node[outerbox,inner xsep=6pt,inner ysep=6pt,minimum width=1.5cm] (RED) [fit = (A) (E)] {};
            \node[below=0.3cm of A,text=red,thick](READ) {$read$};
            \node[below=0.3cm of GRAY,text=gray,thick] {$compute$};
            \node[below=0.3cm of E,text=red,thick](WRITE) {$write$};
        \end{scope}
        \node[codebox,above=\distancelabel of RED] (label) {kernel: hls\_top};
    \end{tikzpicture}
%\end{document}
    \caption{$hls\_top$ is a kernel made of 4 tasks: $T_{1}$ and $T_{2}$ for computing plus two additional tasks $T_{R}$ and $T_{W}$ for reading/writing in order to enable global memory burst transfers.}
    \label{fig:burst-gmem}
\end{figure}
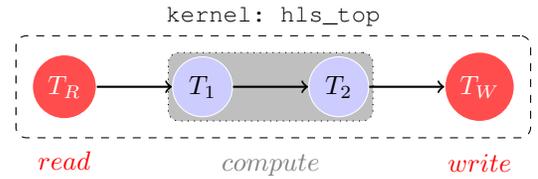

\cref{plot:xil_opt_hls} illustrates how different optimizations dramatically improve the applications' performance in terms of execution time, measured using Vitis analyzer.
In order to observe how task pipelining takes place with dataflow optimization, we launch each kernel 6 times, so that several memory transactions happen consecutively.
In all applications, the DMA engine transfers 25.166 MB from the host to the kernel's global memory.
FLOWER extracts global memory operations from the kernel and automatically generates the dataflow-optimized version, which in turn allows burst transfers for read/write memory.
In fact, in FLOWER, even applications that consist of only a single stage, are divided into at least three tasks which are \emph{read from global memory}, \emph{compute}, and \emph{write to global memory}, enabling pipelined dataflow executions.
\cref{fig:burst-gmem} demonstrate how we utilize global memories.

As described in \cref{sec:flower}, vectorization aggregates several input data (pixels) to vectors, and by replicating the arithmetic operations, processes them at the same time.
Thanks to appropriate packing of data and sequential access patterns on inputs and outputs of the generated kernel, FLOWER enables Vitis to figure out the corresponding memory interface data-width for kernel to global memory transfers.
This data-width is aligned with the size of the vectorized datapath, which in turn results in optimal dataflow transfers throughout the entire design.
Memory controllers of common FPGA cards support a data bus width of up to 512 bits, which in practice defines an upper limit for constructive vectorization.
By combining vectorization and burst transfer optimizations we get the best performance.

Since AnyHLS could not take advantage of the dataflow optimization, it also could not benefit from burst transfers, resulting in execution times that are up to $20\times$ slower.

\subsection{OpenCL}% OpenCL results
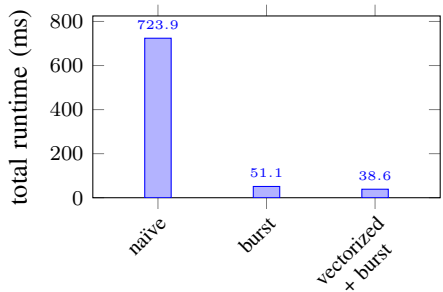
\begin{figure}
    \centering
    \begin{tikzpicture}%[scale = 1]
    \begin{axis}
    [
        small,
        %width=0.30\textwidth,
        %width=5cm,
        width = 0.7\columnwidth,
        height = 4cm,
        ybar=15pt,
        ymin=0,
        ymax=825,
        enlarge x limits=0.3,
        ylabel={total runtime (ms)},
        ylabel near ticks,
        xlabel={},
        symbolic x coords={na\"ive,burst,burst+vectorized}, % these are the specification of coordinates on the x-axis.
        xticklabels={na\"ive,burst,vectorized\\+ burst},
        xticklabel style={align=center},
        xtick=data,
        nodes near coords, % this command is used to mention the y-axis points on the top of the particular bar.
        nodes near coords align={vertical},
        every node near coord/.append style={font=\tiny},
        x tick label style={yshift=0.10cm,rotate=45,anchor=north east}
        ]
    \addplot coordinates { (na\"ive, 723.9) (burst, 51.1) (burst+vectorized, 38.6) };
    \end{axis}
    \end{tikzpicture}
    \vspace{-15pt}
    \caption{Total kernel runtime of the Gaussian filter. The kernel is generated using FLOWER's OpenCL backend for Xilinx FPGAs. The na\"ive version is made of only one kernel/task. The kernel is executed 6 times for a 1024×1024 image on a Xilinx Alveo U280 PCIe accelerator card.}
    \label{plot:xil_ocl_gaussian}
    \end{figure}

\begin{figure}
\centering
\begin{tikzpicture}
    \begin{axis}[
        width = \columnwidth, height = 4cm,
        ybar,
        %ymax=1000,
        %ymin=0,
        %scaled y ticks=base 10:-2,
        ymode=log,
        log basis y={10},
        log ticks with fixed point,
        %max space between ticks=50,
        legend columns = -1, % Horizontal legend
        %legend style={at={(0.5, -0.62)}, anchor = north}, % Place legend below the graph
        legend style={anchor=south east,draw=none,fill=none,at={(current axis.north east)}},
        symbolic x coords = { filter-chain, mean-filter, gaussian, harris, jacobi },
        xtick = { filter-chain, mean-filter, gaussian, harris, jacobi },
        x tick label style={yshift=0.1cm,rotate=45,anchor=north east}, % Rotate horizontal ticks by 45 degrees to save space
        enlarge x limits = 0.08,
        ylabel = {total runtime (ms)},
        bar width = 2mm,
        ]
        \addplot coordinates {(filter-chain, 33.41) (mean-filter, 29.52) (gaussian, 29.5)  (harris, 38.95) (jacobi, 30.31)}; %
        \addplot coordinates {(filter-chain, 24.2)  (mean-filter, 23.57) (gaussian, 21.76) (harris, 28.95) (jacobi, 22.01)}; % _v
        %\addplot coordinates {(filter-chain, 23.5)  (mean-filter, 23.69) }; % _burst_v
        \legend{non-vectorized, vectorized, \begin{comment}vectorized+burst\end{comment}}
    \end{axis}
\end{tikzpicture}
\caption{Evaluation of FLOWER's OpenCL backend on a Bittware 520N-MX card (Intel FPGA). Applications are executed 6 times and total kernel runtime is measured. Image size is $1024\times 1024$.}
\label{plot:intel_ocl}
\end{figure}
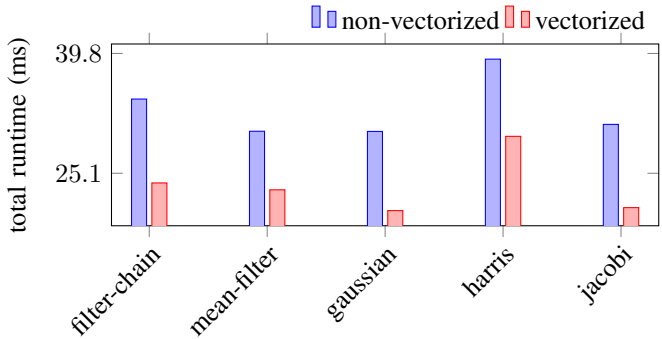

The OpenCL backend of FLOWER generates multi-platform OpenCL device and host codes from the same application.
These codes can be synthesized with both Xilinx Vitis and Intel OpenCL SDK without any modifications.
\cref{plot:xil_ocl_gaussian} and \cref{plot:intel_ocl} show the total execution time of kernels launched 6 times for Xilinx (measured using Vitis analyzer) and Intel OpenCL (measured using the Intel dynamic profiler). %respectively.
While we cannot compare Intel's OpenCL version to the Xilinx one, because of the user licenses of the vendor tools, these charts demonstrate that FLOWER works on both platforms without requiring any change in the application code.

%\subsection{Summary}
%Overall, these results show that FLOWER provides good performance, often beating state-of-the-art approaches thanks to its automatic dataflow optimization.
%At the same time, FLOWER allows programmers to write applications in a single, unified framework that can target different vendor platforms.

\section{Related Work}
FPGA designs applying dataflow optimizations for data access and reuse are significantly more efficient than non-optimized designs:
A climate prediction application runs $800\times$ faster than a na\"ive version when developers correctly address data movement~\cite{9264692,10.1145/3373087.3375296}.
Unfortunately, current C/C++ HLS languages lack the ability to express parallelism and dataflow properly, which means that a huge amount of low-level programming is still required when using those languages.

When it comes to image processing in particular, the literature contains several DSLs and compilers for FPGAs.
They follow various approaches for generating hardware:
SCORE~\cite{10.5555/647927.739401} introduces basic elements of dataflow architectures and is based on TDF which is basically an RTL language.
RIPL~\cite{10.1145/3180481} and Spatial~\cite{10.1145/3192366.3192379} target intermediate languages to generate HDL code, named CAL dataflow~\cite{lucarz:hal-00336520} and Chisel~\cite{6241660}.
Through these intermediate languages, RIPL and Spatial apply basic dataflow optimizations at the HDL level.
The LIFT~\cite{10.1145/3315454.3329957} and Darkroom~\cite{10.1145/2601097.2601174} generate HDL code directly from functional patterns suitable for dataflow programming and do not support data-parallelism.
While these tools construct hardware IPs and interfaces, they do not provide the flexibility of HLS tools that generate low-level RTL through automatic allocation, binding, and scheduling of operators/registers~\cite{5209958}, as in commercial or open-source HLS tools like LegUp~\cite{10.1145/2514740}.

Hipacc~\cite{membarth2016hipacc,reiche2017hipaccfpga}, HeteroHalide~\cite{10.1145/3373087.3375320}, PolyMage~\cite{7756767}, the Merlin compiler~\cite{10.1145/2934583.2953984}, and AnyHLS~\cite{oezkan2020anyhls} target C/C++ codes in their backend while applying optimization passes suitable for HLS tools such as Xilinx Vitis or Intel HLS compilers.
AnyHLS makes use of the AnyDSL compiler~\cite{7054200} to partially evaluate~\cite{Futamura1999,leissa2018anydsl} higher-order functions in order to generate optimized OpenCL/C++ HLS codes.
Dahlia~\cite{10.1145/3385412.3385974} is similar to AnyHLS, in that it provides a general purpose language for generating specialized HLS C++ code with performance predictability.
Dahlia uses a rich type system to deliver performance, while AnyHLS relies on function specialization.

There are several studies presenting the importance of dataflow designs for stream processing, most notably OxiGen~\cite{8425390}, which relies on the MaxCompiler, or Frost~\cite{8445108}, which uses the SDAccel synthesis tool (now superseded by Xilinx Vitis) and doesn't strictly follow the canonical form required for Vitis.
Both tools require ad-hoc host programs to be written by the user.
HLS dataflow transformations have also demonstrated benefits in packet processing and deep learning pipelines~\cite{8735559,DBLP:journals/corr/ChengW16,10.1145/3020078.3021744}.

In contrast to FLOWER, most of these frameworks, including Hipacc and PolyMage, generate a dataflow design through C++ template metaprogramming, and their approach does not rigorously follow the canonical form recommended by Xilinx.
They only support Vivado HLS and are not updated for the new Vitis toolchain from Xilinx.
Additionally, most of these studies only evaluate the generated C/C++ IPs on Zynq SOC platforms, which are only designed for embedded systems.
StencilFlow~\cite{DBLP:journals/corr/abs-2010-15218}, and other frameworks~\cite{10.1145/3174243.3174248,8884961} show how an optimized Intel OpenCL implementation is beneficial for dataflow and stream processing designs.
AFFIX~\cite{10.1145/3289602.3293907} also provides a scalable OpenCL library for vision algorithms via Intel OpenCL on FPGAs.
While our work is based on stateless synchronous dataflow graphs, others~\cite{10.1145/3295500.3356173} introduce a data-centric model for stateful dataflow multigraphs, which relies on SDAccel and Xilinx OpenCL code.

\section{Conclusion}
Dataflow transformations, low-level optimizations, and host code development are essential building blocks to achieve an efficient design for streaming applications, as shown by the examples in this paper.
FLOWER allows programmers to write their applications in a high-level library, and automatically introduces the required transformations and optimizations.
Our results demonstrate that our work is not only faster compared to similar tools, but also increases productivity, in contrast with alternatives where manual work is required to do those transformations.
FLOWER is fully compatible with both Xilinx and Intel FPGA accelerator cards, allowing to use the same code to drive two different devices.
In the future, we would like to take advantage of the LLVM IR backend of AnyDSL to target the recently open-sourced front-end of Vitis, in order to perform even more optimizations.

\bibliographystyle{IEEEtran}
\bibliography{references}

\begin{thebibliography}{10}
\providecommand{\url}[1]{#1}
\csname url@samestyle\endcsname
\providecommand{\newblock}{\relax}
\providecommand{\bibinfo}[2]{#2}
\providecommand{\BIBentrySTDinterwordspacing}{\spaceskip=0pt\relax}
\providecommand{\BIBentryALTinterwordstretchfactor}{4}
\providecommand{\BIBentryALTinterwordspacing}{\spaceskip=\fontdimen2\font plus
\BIBentryALTinterwordstretchfactor\fontdimen3\font minus
  \fontdimen4\font\relax}
\providecommand{\BIBforeignlanguage}[2]{{%
\expandafter\ifx\csname l@#1\endcsname\relax
\typeout{** WARNING: IEEEtran.bst: No hyphenation pattern has been}%
\typeout{** loaded for the language `#1'. Using the pattern for}%
\typeout{** the default language instead.}%
\else
\language=\csname l@#1\endcsname
\fi
#2}}
\providecommand{\BIBdecl}{\relax}
\BIBdecl

\bibitem{9264692}
J.~de~Fine~Licht, M.~Besta, S.~Meierhans, and T.~Hoefler, ``Transformations of
  high-level synthesis codes for high-performance computing,'' \emph{IEEE
  Transactions on Parallel \& Distributed Systems}, vol.~32, no.~05, pp.
  1014--1029, Jan. 2021.

\bibitem{8891819}
T.~{Kenter}, ``Invited tutorial: {OpenCL} design flows for intel and xilinx
  {FPGAs}: Using common design patterns and dealing with vendor-specific
  differences,'' in \emph{Sixth International Workshop on FPGAs for Software
  Programmers (FSP Workshop)}, 2019, pp. 1--8.

\bibitem{8945776}
N.~{Brown} and D.~{Dolman}, ``It's all about data movement: Optimising {FPGA}
  data access to boost performance,'' in \emph{IEEE/ACM International Workshop
  on Heterogeneous High-performance Reconfigurable Computing (H2RC)}, 2019, pp.
  1--10.

\bibitem{oezkan2020anyhls}
M.~A. Özkan, A.~Pérard-Gayot, R.~Membarth, P.~Slusallek, R.~Leißa, S.~Hack,
  J.~Teich, and F.~Hannig, ``{AnyHLS}: High-level synthesis with partial
  evaluation,'' \emph{IEEE Transactions on Computer-Aided Design of Integrated
  Circuits and Systems (TCAD) (Proceedings of CODES+ISSS 2020)}, vol.~39,
  no.~11, pp. 3202--3214, Sep. 2020.

\bibitem{leissa2018anydsl}
R.~Leißa, K.~Boesche, S.~Hack, A.~Pérard-Gayot, R.~Membarth, P.~Slusallek,
  A.~Müller, and B.~Schmidt, ``{AnyDSL}: A partial evaluation framework for
  programming high-performance libraries,'' \emph{Proceedings of the ACM on
  Programming Languages (PACMPL)}, vol.~2, no. OOPSLA, pp. 119:1--119:30, Nov.
  2018, {HiPEAC 2018 Paper Award}.

\bibitem{DBLP:journals/corr/abs-1807-01340}
J.~Cong, Z.~Fang, Y.~Hao, P.~Wei, C.~H. Yu, C.~Zhang, and P.~Zhou,
  ``Best-effort {FPGA} programming: {A} few steps can go a long way,''
  \emph{CoRR}, vol. abs/1807.01340, 2018.

\bibitem{10.1145/3373087.3375320}
J.~Li, Y.~Chi, and J.~Cong, ``{HeteroHalide}: From image processing {DSL} to
  efficient {FPGA} acceleration,'' in \emph{Proceedings of the 2020 ACM/SIGDA
  International Symposium on Field-Programmable Gate Arrays}.\hskip 1em plus
  0.5em minus 0.4em\relax ACM, 2020, pp. 51--57.

\bibitem{7756767}
N.~{Chugh}, V.~{Vasista}, S.~{Purini}, and U.~{Bondhugula}, ``A {DSL} compiler
  for accelerating image processing pipelines on {FPGAs},'' in
  \emph{International Conference on Parallel Architecture and Compilation
  Techniques (PACT)}, 2016, pp. 327--338.

\bibitem{10.1145/3377555.3377899}
S.~Purini, V.~Benara, Z.~Choudhury, and U.~Bondhugula, ``Bitwidth customization
  in image processing pipelines using interval analysis and smt solvers,'' in
  \emph{Proceedings of the 29th International Conference on Compiler
  Construction}.\hskip 1em plus 0.5em minus 0.4em\relax ACM, 2020, pp.
  167--178.

\bibitem{10.1145/2897824.2925892}
J.~Hegarty, R.~Daly, Z.~DeVito, J.~Ragan-Kelley, M.~Horowitz, and P.~Hanrahan,
  ``Rigel: Flexible multi-rate image processing hardware,'' \emph{ACM Trans.
  Graph.}, vol.~35, no.~4, Jul. 2016.

\bibitem{membarth2016hipacc}
R.~Membarth, O.~Reiche, F.~Hannig, J.~Teich, M.~Körner, and W.~Eckert,
  ``{Hipacc: A Domain-Specific Language and Compiler for Image Processing},''
  \emph{Transactions on Parallel and Distributed Systems (TPDS)}, vol.~27,
  no.~1, pp. 210--224, Jan. 2016.

\bibitem{reiche2017hipaccfpga}
O.~Reiche, M.~A. Özkan, R.~Membarth, J.~Teich, and F.~Hannig, ``{Generating
  FPGA-based Image Processing Accelerators with Hipacc},'' in \emph{Proceedings
  of the International Conference On Computer Aided Design (ICCAD)}.\hskip 1em
  plus 0.5em minus 0.4em\relax Irvine, CA, USA: IEEE, Nov. 2017, pp.
  1026--1033, {Invited Paper}.

\bibitem{10.1145/3373087.3375296}
J.~de~Fine~Licht, G.~Kwasniewski, and T.~Hoefler, ``Flexible communication
  avoiding matrix multiplication on {FPGA} with high-level synthesis,'' in
  \emph{Proceedings of the 2020 ACM/SIGDA International Symposium on
  Field-Programmable Gate Arrays}.\hskip 1em plus 0.5em minus 0.4em\relax ACM,
  2020, pp. 244--254.

\bibitem{10.5555/647927.739401}
E.~Caspi, M.~Chu, R.~Huang, J.~Yeh, J.~Wawrzynek, and A.~DeHon, ``Stream
  computations organized for reconfigurable execution (score),'' in \emph{10th
  International Workshop on Field-Programmable Logic and Applications}.\hskip
  1em plus 0.5em minus 0.4em\relax Springer, 2000, pp. 605--614.

\bibitem{10.1145/3180481}
R.~Stewart, K.~Duncan, G.~Michaelson, P.~Garcia, D.~Bhowmik, and A.~Wallace,
  ``{RIPL}: A parallel image processing language for {FPGAs},'' \emph{ACM
  Trans. Reconfigurable Technol. Syst.}, vol.~11, no.~1, Mar. 2018.

\bibitem{10.1145/3192366.3192379}
D.~Koeplinger, M.~Feldman, R.~Prabhakar, Y.~Zhang, S.~Hadjis, R.~Fiszel,
  T.~Zhao, L.~Nardi, A.~Pedram, C.~Kozyrakis, and K.~Olukotun, ``Spatial: A
  language and compiler for application accelerators,'' in \emph{Proceedings of
  the 39th ACM SIGPLAN Conference on Programming Language Design and
  Implementation}.\hskip 1em plus 0.5em minus 0.4em\relax ACM, 2018, pp.
  296--311.

\bibitem{lucarz:hal-00336520}
C.~Lucarz, M.~Mattavelli, M.~Wipliez, G.~Roquier, M.~Raulet, J.~W. Janneck,
  I.~D. Miller, and D.~B. Parlour, ``{Dataflow/Actor-Oriented language for the
  design of complex signal processing systems},'' in \emph{{Conference on
  Design and Architectures for Signal and Image Processing (DASIP 2008)}},
  Bruxelles, Belgium, Nov. 2008, pp. 1--8.

\bibitem{6241660}
J.~{Bachrach}, H.~{Vo}, B.~{Richards}, Y.~{Lee}, A.~{Waterman},
  R.~{Avižienis}, J.~{Wawrzynek}, and K.~{Asanović}, ``Chisel: Constructing
  hardware in a scala embedded language,'' in \emph{Design Automation
  Conference}, 2012, pp. 1212--1221.

\bibitem{10.1145/3315454.3329957}
M.~Kristien, B.~Bodin, M.~Steuwer, and C.~Dubach, ``High-level synthesis of
  functional patterns with lift,'' in \emph{Proceedings of the 6th ACM SIGPLAN
  International Workshop on Libraries, Languages and Compilers for Array
  Programming}.\hskip 1em plus 0.5em minus 0.4em\relax ACM, 2019, pp. 35--45.

\bibitem{10.1145/2601097.2601174}
J.~Hegarty, J.~Brunhaver, Z.~DeVito, J.~Ragan-Kelley, N.~Cohen, S.~Bell,
  A.~Vasilyev, M.~Horowitz, and P.~Hanrahan, ``Darkroom: Compiling high-level
  image processing code into hardware pipelines,'' \emph{ACM Trans. Graph.},
  vol.~33, no.~4, Jul. 2014.

\bibitem{5209958}
P.~{Coussy}, D.~D. {Gajski}, M.~{Meredith}, and A.~{Takach}, ``An introduction
  to high-level synthesis,'' \emph{IEEE Design Test of Computers}, vol.~26,
  no.~4, pp. 8--17, 2009.

\bibitem{10.1145/2514740}
A.~Canis, J.~Choi, M.~Aldham, V.~Zhang, A.~Kammoona, T.~Czajkowski, S.~D.
  Brown, and J.~H. Anderson, ``{LegUp}: An open-source high-level synthesis
  tool for {FPGA}-based processor/accelerator systems,'' \emph{ACM Trans.
  Embed. Comput. Syst.}, vol.~13, no.~2, Sep. 2013.

\bibitem{10.1145/2934583.2953984}
J.~Cong, M.~Huang, P.~Pan, D.~Wu, and P.~Zhang, ``Software infrastructure for
  enabling {FPGA}-based accelerations in data centers: Invited paper,'' in
  \emph{Proceedings of the 2016 International Symposium on Low Power
  Electronics and Design}.\hskip 1em plus 0.5em minus 0.4em\relax ACM, 2016,
  pp. 154--155.

\bibitem{7054200}
R.~{Leißa}, M.~{Köster}, and S.~{Hack}, ``A graph-based higher-order
  intermediate representation,'' in \emph{IEEE/ACM International Symposium on
  Code Generation and Optimization (CGO)}, Feb. 2015, pp. 202--212.

\bibitem{Futamura1999}
Y.~Futamura, ``Partial evaluation of computation process--an approach to a
  compiler-compiler,'' \emph{Higher-Order and Symbolic Computation}, vol.~12,
  no.~4, pp. 381--391, Dec. 1999.

\bibitem{10.1145/3385412.3385974}
R.~Nigam, S.~Atapattu, S.~Thomas, Z.~Li, T.~Bauer, Y.~Ye, A.~Koti, A.~Sampson,
  and Z.~Zhang, ``Predictable accelerator design with time-sensitive affine
  types,'' in \emph{Proceedings of the 41st ACM SIGPLAN Conference on
  Programming Language Design and Implementation}.\hskip 1em plus 0.5em minus
  0.4em\relax ACM, 2020, pp. 393--407.

\bibitem{8425390}
F.~{Peverelli}, M.~{Rabozzi}, E.~{Del Sozzo}, and M.~D. {Santambrogio},
  ``Oxigen: A tool for automatic acceleration of c functions into dataflow
  {FPGA}-based kernels,'' in \emph{IEEE International Parallel and Distributed
  Processing Symposium Workshops (IPDPSW)}, 2018, pp. 91--98.

\bibitem{8445108}
E.~D. Sozzo, R.~Baghdadi, S.~Amarasinghe, and M.~D. Santambrogio, ``A unified
  backend for targeting {FPGAs} from {DSLs},'' in \emph{IEEE 29th International
  Conference on Application-specific Systems, Architectures and Processors
  (ASAP)}.\hskip 1em plus 0.5em minus 0.4em\relax Los Alamitos, CA, USA: IEEE
  Computer Society, Jul. 2018, pp. 1--8.

\bibitem{8735559}
H.~{Eran}, L.~{Zeno}, Z.~{István}, and M.~{Silberstein}, ``Design patterns for
  code reuse in hls packet processing pipelines,'' in \emph{IEEE 27th Annual
  International Symposium on Field-Programmable Custom Computing Machines
  (FCCM)}, 2019, pp. 208--217.

\bibitem{DBLP:journals/corr/ChengW16}
S.~Cheng and J.~Wawrzynek, ``High level synthesis with a dataflow architectural
  template,'' \emph{CoRR}, vol. abs/1606.06451, 2016.

\bibitem{10.1145/3020078.3021744}
Y.~Umuroglu, N.~J. Fraser, G.~Gambardella, M.~Blott, P.~Leong, M.~Jahre, and
  K.~Vissers, ``Finn: A framework for fast, scalable binarized neural network
  inference,'' in \emph{Proceedings of the 2017 ACM/SIGDA International
  Symposium on Field-Programmable Gate Arrays}.\hskip 1em plus 0.5em minus
  0.4em\relax ACM, 2017, pp. 65--74.

\bibitem{DBLP:journals/corr/abs-2010-15218}
J.~de~Fine~Licht, A.~Kuster, T.~D. Matteis, T.~Ben{-}Nun, D.~Hofer, and
  T.~Hoefler, ``{StencilFlow}: Mapping large stencil programs to distributed
  spatial computing systems,'' \emph{CoRR}, vol. abs/2010.15218, 2020.

\bibitem{10.1145/3174243.3174248}
H.~R. Zohouri, A.~Podobas, and S.~Matsuoka, ``Combined spatial and temporal
  blocking for high-performance stencil computation on {FPGAs} using
  {OpenCL},'' in \emph{Proceedings of the 2018 ACM/SIGDA International
  Symposium on Field-Programmable Gate Arrays}.\hskip 1em plus 0.5em minus
  0.4em\relax ACM, 2018, pp. 153--162.

\bibitem{8884961}
S.~{Wu}, D.~{Hu}, S.~{Ibrahim}, H.~{Jin}, J.~{Xiao}, F.~{Chen}, and H.~{Liu},
  ``When {FPGA}-accelerator meets stream data processing in the edge,'' in
  \emph{IEEE 39th International Conference on Distributed Computing Systems
  (ICDCS)}, 2019, pp. 1818--1829.

\bibitem{10.1145/3289602.3293907}
S.~Taheri, P.~Behnam, E.~Bozorgzadeh, A.~Veidenbaum, and A.~Nicolau, ``{AFFIX}:
  Automatic acceleration framework for {FPGA} implementation of {OpenVX} vision
  algorithms,'' in \emph{Proceedings of the 2019 ACM/SIGDA International
  Symposium on Field-Programmable Gate Arrays}.\hskip 1em plus 0.5em minus
  0.4em\relax ACM, 2019, pp. 252--261.

\bibitem{10.1145/3295500.3356173}
T.~Ben-Nun, J.~de~Fine~Licht, A.~N. Ziogas, T.~Schneider, and T.~Hoefler,
  ``Stateful dataflow multigraphs: A data-centric model for performance
  portability on heterogeneous architectures,'' in \emph{Proceedings of the
  International Conference for High Performance Computing, Networking, Storage
  and Analysis}.\hskip 1em plus 0.5em minus 0.4em\relax ACM, 2019.

\end{thebibliography}
\end{document}